\documentstyle[12pt]{article}

\advance\textheight by \topskip
\begin{document}
%
\thispagestyle{empty}
\setcounter{page}{0}

\newcommand{\be}{\begin{equation}}
\newcommand{\ee}{\end{equation}}
\newcommand{\br}{\begin{eqnarray}}
\newcommand{\er}{\end{eqnarray}}
\newcommand{\ba}{\begin{array}}
\newcommand{\ea}{\end{array}}
\newcommand{\bi}{\begin{itemize}}
\newcommand{\ei}{\end{itemize}}
\newcommand{\bc}{\begin{center}}
\newcommand{\ec}{\end{center}}
\newcommand{\ul}{\underline}
\newcommand{\ol}{\overline}
\newcommand{\eebbww}{$e^+e^-\rightarrow b\bar b W^+W^-$}
\newcommand{\bb}{$ b\bar b \ $}
\newcommand{\ttb}{$ t\bar t \ $}
\newcommand{\ar}{\rightarrow}
\newcommand{\mssm}{${\cal {MSSM}}\ $}
\newcommand{\sm}{${\cal {SM}}\ $}
\newcommand{\Dir}{\kern -6.4pt\Big{/}}
\newcommand{\Dirin}{\kern -10.4pt\Big{/}\kern 4.4pt}
\newcommand{\DDir}{\kern -7.6pt\Big{/}}
\newcommand{\DGir}{\kern -6.0pt\Big{/}}

\def\Ord{\buildrel{\scriptscriptstyle <}\over{\scriptscriptstyle\sim}}
\def\OOrd{\buildrel{\scriptscriptstyle >}\over{\scriptscriptstyle\sim}}
\def\pl #1 #2 #3 {{\it Phys.~Lett.} {\bf#1} (#2) #3}
\def\np #1 #2 #3 {{\it Nucl.~Phys.} {\bf#1} (#2) #3}
\def\zp #1 #2 #3 {{\it Z.~Phys.} {\bf#1} (#2) #3}
\def\pr #1 #2 #3 {{\it Phys.~Rev.} {\bf#1} (#2) #3}
\def\prep #1 #2 #3 {{\it Phys.~Rep.} {\bf#1} (#2) #3}
\def\prl #1 #2 #3 {{\it Phys.~Rev.~Lett.} {\bf#1} (#2) #3}
\def\mpl #1 #2 #3 {{\it Mod.~Phys.~Lett.} {\bf#1} (#2) #3}
\def\rmp #1 #2 #3 {{\it Rev. Mod. Phys.} {\bf#1} (#2) #3}
\def\xx #1 #2 #3 {{\bf#1}, (#2) #3}
\def\preprint{{\it preprint}}

\begin{flushright}
{\large DFTT 60/94}\\
{\large DTP/94/104}\\
{\large MZ-THEP-95-10}\\
{\rm January 1995\hspace*{.5 truecm}}\\
\end{flushright}

\vspace*{\fill}

\begin{center}
{\Large \bf QCD effects and $b$-tagging at LEP I\footnote{Work
supported in part by Ministero
dell' Universit\`a e della Ricerca Scientifica.\\
$\dag$ E-mail: Stefano.Moretti@durham.ac.uk,
tausk@vipmzw.physik.uni-mainz.de}}\\[2.cm]
{\large S.~Moretti$^{a,b,\dag}$ and J.B.~Tausk$^{c,\dag}$ }\\[2cm]
{\it a) Department of Physics, University of Durham,}\\
{\it    South Road, Durham DH1 3LE, U.K.}\\[1cm]
{\it b) Dipartimento di Fisica Teorica, Universit\`a di Torino,}\\
{\it    and INFN, Sezione di Torino,}\\
{\it    V. Pietro Giuria 1, 10125 Torino, Italy.}\\[1cm]
{\it c) Johannes Gutenberg-Universit\"at,
        Institut f\"ur Physik (THEP),}\\
{\it    Staudingerweg 7,
        D-55099 Mainz, Germany.}\\[2cm]
\end{center}

\vspace*{\fill}

\begin{abstract}
{\normalsize
\noindent
We analyze the impact of using $b$-tagged samples in studying non-Abelian
effects due to QCD in $e^+e^-\ar 4$jet events at $\sqrt s=M_{Z^0}$, using
angular variable analyses and comparisons with $e^+e^-\ar 3 \mbox{jet}\gamma$
events. We find that QCD effects are largely enhanced in $b$-quark samples
with respect to `unflavoured' ones, where energy-ordering is used to
distinguish between gluon and quark jets. We show that the $b$-quark mass
influences the angular distributions significantly and should not be
neglected.}
\end{abstract}

\vspace*{\fill}

\newpage

\section{Introduction}

In recent years the experiments ALEPH, DELPHI, L3 and OPAL at LEP I
have performed a number of measurements of the process $e^+e^-\ar Z^0 \ar
hadrons$ in order to point out effects due to QCD \cite{Hebbeker}.
Important results have been achieved.
The strong coupling constant $\alpha_s$ has been determined from jet
rates and from shape variables \cite{alphas} and both the
flavour independence \cite{flavour} and the running with $\sqrt s$
\cite{running} have been verified.
Three- \cite{3jd} and four-jet \cite{4jd} distributions have been studied
and their behaviour agrees with
QCD predictions calculated to second order in $\alpha_s$.
The colour factors, which determine the gauge
group responsible for strong interactions, have been measured
\cite{colfac}. Abelian models alternative to QCD
have been ruled out and the coupling of the QCD triple gluon vertex has been
verified to be in agreement with QCD predictions \cite{Abelian}.

Concerning the latter tests, several variables sensitive to differences
between QCD and Abelian models have been proposed
\cite{EK,NR,NRmod,BZ,KSW}. One of the main differences is the predicted
relative contribution of $e^+e^-\ar q\bar q q'\bar q'$ events in $e^+e^-\ar
4$jet samples: about 5\% in QCD but 30\% in Abelian models \cite{nonAbel}.
The tests are preferentially based on angular correlations between jets,
which are usually ordered in energy and where the two most energetic jets
are ``identified'' with the primary quarks (i.e., from the $Z^0$ decay).
Among these angular variables, some of the most widely used in experimental
analyses are the modified Nachtmann-Reiter angle $\theta_{NR}^*$
\cite{NR,NRmod}, the Bengtsson-Zerwas angle $\chi_{BZ}$ \cite{BZ}, the
K\"orner-Schierholz-Willrodt angle $\Phi_{KSW}$ \cite{KSW} and the angle
between the two least energetic jets $\theta_{34}$. The distributions in
these variables are quite different for $e^+e^-\to q\bar{q}gg$ and
$e^+e^-\to q\bar{q}q'\bar{q'}$ events.

Another way of searching for evidence of effects due to QCD, and only
partially exploited so far, is to use photon samples. In particular, in
order to isolate the triple gluon vertex contribution, one can compare
$4$jet- and $3 \mbox{jet}\gamma$-samples. This approach is based on the
similarity of photon and gluon bremsstrahlung off quarks \cite{workshop}
and has already been adopted in ref.~\cite{Mattig}. However, there is also
an obvious difference between photons and gluons, which is that only
photons can also be radiated off the initial state electrons and positrons.
Therefore some care is needed to eliminate the distortion due to this
Initial State Radiation (ISR) from the $3 \mbox{jet}\gamma$-sample before a
direct comparison with the 4jet sample can be made.

It is the purpose of this paper to study to what extent the techniques of
flavour identification, that are rapidly being developed by the various
experimental collaborations at LEP I \cite{btagg}, turn out to be useful in
recognizing effects of QCD from the analysis of $b$-tagged $4$jet- and
$3 \mbox{jet}\gamma$-samples, following both the approaches described above.
The most widely used methods to recognize $b$-quark jets are
probably the following:
\begin{itemize}
\item{reconstruction of semileptonic $b$-decays by observing a high
		     $p_T$ lepton;}
\item{lifetime tagging by detecting a secondary vertex;}
\item{reconstruction using kinematical ``event shape'' or
		   ``jet shape'' variables.}
\end{itemize}
\noindent
Their main features are summarized in refs.~\cite{Squarcia,deangelis}.
Recently, in addition to these and other conventional methods
\cite{conventional} also the identification of gluon and quark jets by
means of neural networks has been proposed \cite{neural}.

In our opinion, there is an important motivation for this study. As many
authors have pointed out, the angular variables described here are most
useful for emphasizing the non-Abelian features of QCD if one distinguishes
between quark and gluon jets and assigns the four-momenta of the final
states to the corresponding particles. If that is not possible, the best
one can do is to order the jets in energy\footnote{In $Z^0\ar q\bar qgg$
events the lowest energy parton is a gluon in $\approx84\%$ of cases,
whereas the percentage of events in which the two lowest energy partons are
both gluons is only $\approx53\%$ \cite{nonAbel}.}. However, with
$b$-tagging, it is at least possible to distinguish some quark jets, namely
those originating from $b$-quarks, from gluon jets. This opens the prospect
of observing the non-Abelian structure of QCD much more clearly by
selecting events containing $b$-jets. Since this means that all final
states contain at least two $b$-quarks, the effect of the the $b$-quark
mass becomes much more important than it was before, and it must be
properly taken into account in the analysis. The required matrix elements
have only recently become available \cite{noiPL,noiNP,noiPr}\footnote{An
updated review on matrix element computations for multi-jet production in
$e^+ e^-$ reactions presented in the literature can be found, e.g., in the
introduction of ref.~\cite{noiNP}.}.

The plan of the paper is as follows. In section~\ref{subs:calc} we give
details of the calculation, of the algorithms used in the phenomenological
analysis and the numerical values adopted for the various parameters.
Section~\ref{subs:difdis} is devoted to a discussion of the differential
distributions in four angular variables, and section~\ref{subs:photon} to
the possibility of using $3 \mbox{jet}\gamma$ samples. In
section~\ref{subs:me} we study the sensitivity of our results to the
$b$-quark mass, and we draw our conclusions in section~\ref{subs:concs}.

\section{Calculation}
\label{subs:calc}

The Feynman diagrams describing at tree-level the reactions
\be\label{proc1}
e^+ + e^-\ar q_1 + \bar q_2 + {q'}_3 + \bar {q'}_4,
\ee
\be\label{proc2}
e^+ + e^-\ar q_1 + \bar q_2 + g_3 + g_4,
\ee
\be\label{proc3}
e^+ + e^-\ar q_1 + \bar q_2 + g_3 + \gamma_4,
\ee
where $q^{(')}=u,d,s,c$ and $b$,
are shown in figs.~1-2. In the present analysis
we have computed the corresponding matrix elements
with the same {\tt FORTRAN} code already used in
refs.~\cite{noiPL,noiNP,noiPr}, which takes all masses
and both the $\gamma$ and $Z^0$ intermediate contributions
into account exactly. For all the details of the computation,
as well as for the explicit helicity amplitude formulae, we refer to
\cite{noiNP}.

The matrix element squared $|{\cal M}|^2$ for the four quark
process~(\ref{proc1}) can be written as
\be
\label{M2-4q}
|{\cal M}|^2=
 C_+^{q\bar{q}q'\bar{q}'} |{\cal M}_+|^2 +
 C_-^{q\bar{q}q'\bar{q}'} |{\cal M}_-|^2,
\ee
with
\be
 C_{\pm}^{q\bar{q}q'\bar{q}'} =
 \frac{1}{2} N_C C_F ( T_F \mp (C_F - \frac{1}{2} C_A) ),
\ee
\be
{\cal M}_{\pm} =  {\cal M}_1 + {\cal M}_2 + {\cal M}_3 + {\cal M}_4
 \pm \delta^{qq'}({\cal M}_5 + {\cal M}_6 + {\cal M}_7 + {\cal M}_8),
\ee
where ${\cal M}_i$ is the amplitude corresponding to the
$i$-th diagram in fig.~1,
$N_C=3$ the number of colours,  $C_F=(N_C^2-1)/(2N_C)=4/3$ and $C_A=N_C$
the Casimir operators of the fundamental and adjoint representations of
the gauge group $SU(N_C)$, and $T_F=1/2$ the normalisation of the
generators of the fundamental representation.

The matrix element squared for the two quark and two gluon process
(\ref{proc2}), can be split into two gauge invariant parts as
follows \cite{HagiZepp}:
\be
\label{M2-2q2g}
|{\cal M}|^2=
 C_a^{q\bar qgg}|{{\cal M}_a}|^2+C_b^{q\bar qgg}|{{\cal M}_b}|^2,
\ee
where
\be
{{\cal M}_a}=\sum_{i=1,6}{{\cal M}_i},
\ee
\be
{{\cal M}_b}={{\cal M}_1}+{{\cal M}_3}+{{\cal M}_5}
	     -{{\cal M}_2}-{{\cal M}_4}-{{\cal M}_6}
	     -2 \mbox{ i}[{{\cal M}_7}+{{\cal M}_8}],
\ee
${{\cal M}_i}$, $i=1,...8$, corresponds to the $i$-th diagram in
fig.~2, and
\be
C_a^{q\bar qgg}=\frac{1}{2}N_C C_F(2C_F-\frac{1}{2}C_A)=\frac{7}{3},
\quad\quad\quad\quad
C_b^{q\bar qgg}=\frac{1}{2}N_C C_F\frac{1}{2}C_A =3.
\ee
The second term in eq.~(\ref{M2-2q2g}) is characteristic of non-Abelian
theories and would be absent in any Abelian model.

Finally, for the production of two quarks, a gluon and a photon (\ref{proc3}),
the matrix element squared is (up to a constant factor, see later on):
\be
\label{M2-2qgph}
|{\cal M}|^2 =
 C^{q\bar qg\gamma}|e_q{\cal M}_a + e_e {\cal M}_{ISR}|^2,
\ee
with
\be
C^{q\bar qg\gamma} = N_C C_F,
\ee
where
${\cal M}_a$ is the sum of the diagrams of fig.~2a,
${\cal M}_{ISR}$ the sum of the diagrams of fig.~2c,
and $e_e$ and $e_q$ are the electric charges of the electron $e$ and
of the quark $q$ in the final state.

The Abelian model we compare with QCD
is the one introduced in \cite{qed}\footnote{Of course, we know
that such a model has been already ruled out in other contexts,
e.g., by measurements of the energy dependence
of multi-jet production rates in $e^+e^-$ annihilation \cite{ruledout},
but we regard it mainly as a useful tool to demonstrate the
sensitivity of the introduced angular variables to the various
features of QCD.}.
Here gluons have no colour and no self-coupling: therefore, only
diagrams of fig.~1 and fig.~2a survive. The cross sections of
processes (\ref{proc1})-(\ref{proc2}) for the Abelian case can be obtained
from the QCD ones by simply replacing the group constants of QCD by
those appropriate for the Abelian model: i.e.,
$C_A=3\ar 0$, $C_F=4/3\ar 1$ and $T_F=1/2 \ar 3$.
The ``Abelian coupling constant'' $\alpha_A$ is fixed
to $(4/3)\alpha_s$, so that the ratio of the two-jet and three-jet cross
sections agrees with experiment.

We have analyzed the processes (\ref{proc1})-(\ref{proc3}) adopting
four different jet-finding algorithms. They are identified through
their clustering variable $y_{ij}$. They are the JADE scheme (J) \cite{JADE}
based on the variable
\be\label{JADE}
y^J_{ij} = {{2E_i E_j(1-\cos\theta_{ij})}\over{s}},
\ee
and its ``E'' variation (E)\footnote{At lowest order, the E and JADE
schemes are equivalent for massless particles.}
\be\label{E}
y^E_{ij} = \frac{(p_i+p_j)\cdot(p_i+p_j)}{s},
\ee
the Durham scheme (D) \cite{DURHAM}
\be\label{DURHAM}
y^D_{ij} = {{2\min (E^2_i, E^2_j)(1-\cos\theta_{ij})}
\over{s}},
\ee
and the Geneva algorithm (G) \cite{GENEVA}
\be\label{GENEVA}
y^G_{ij} = \frac{8}{9} {{E_iE_j(1-\cos\theta_{ij})}\over{(E_i+E_j)^2}}.
\ee
For all of them the two (pseudo)particles $i$ and $j$
(with energy $E_i$ and $E_j$, respectively) for which $y_{ij}$
is minimum are combined into a single pseudoparticle $k$ of momentum
$P_k$ given by the formula
\be
P_k=P_i+P_j.
\ee
The procedure is iterated until all pseudoparticle pairs satisfy $y_{ij}\ge
y_{cut}$. The various characteristics of these algorithms are well
summarized in ref.~\cite{GENEVA}. In our lowest order calculation,
the four jet cross section for a given algorithm is simply equal to
the four parton cross section with a cut $y_{ij}\ge y_{cut}$ on all
pairs of partons $(i,j)$.

Concerning the numerical part of our work, we have taken
$\alpha_{em}= 1/128$ and  $\sin^2\theta_W\equiv s^2_W=0.23$, while
for the $Z^0$ boson mass and width we have adopted the values
$M_{Z^0}=91.1$ GeV and $\Gamma_{Z^0}=2.5$ GeV, respectively.
For the quarks we have: $m_c=1.7$ GeV and $m_b=5.0$ GeV while
the flavours $u$, $d$ and $s$ have been considered massless.
Finally, the strong coupling constant  $\alpha_s$ has been set equal to
$0.115$.

\section{Angular variables}
\label{subs:difdis}

We study the following four variables: the modified Nachtmann-Reiter
angle, $\theta_{NR}^*$, the Beng\-ts\-son-Zerwas angle, $\chi_{BZ}$, a
modification of the K\"orner-Schierholz-Willrodt angle we denote by
$\Phi_{KSW}^*$, and the angle between jets 3 and 4, $\theta_{34}$, in two
different situations:
(a) with $b$-tagging and
(b) without $b$-tagging.

In case (a), we consider four-jet events where two of the
jets contain a $b$ or a $\bar{b}$. Let us call them jet 1
and jet 2. It does not matter which one is the $b$- and which
the $\bar{b}$-jet.
We do not make any assumptions about jets 3
and 4; they may be either gluon or quark jets, or even
$b$-jets. Then, in terms of the three-momenta
$\vec{p}_1,\dots,{\vec p}_4$ of jets $1,\ldots,4$, the angles
$\theta_{NR}^*$, $\chi_{BZ}$ and $\theta_{34}$ are defined by
\be
\label{def:NR}
\theta_{NR}^*=\angle({\vec p}_1-{\vec p}_2,{\vec p}_3-{\vec p}_4),
\ee
\be
\label{def:BZ}
\chi_{BZ}=\angle({\vec p}_1 \times {\vec p}_2,{\vec p}_3 \times {\vec p}_4),
\ee
and
\be
\label{def:34}
\theta_{34}=\angle({\vec p}_3,{\vec p}_4).
\ee
For events where
\be
\label{def:KSW1}
|{\vec p}_1+{\vec p}_3| > |{\vec p}_1+{\vec p}_4|
\ee
we define
\be
\label{def:KSW2}
\Phi_{KSW}^*=
\angle({\vec p}_1\times{\vec p}_3,{\vec p}_2\times{\vec p}_4).
\ee
In the opposite case, we define $\Phi_{KSW}^*$ with ${\vec p}_3$ and
${\vec p}_4$ interchanged. The definition in
eqs.~(\ref{def:KSW1})-(\ref{def:KSW2}) is equivalent to the
original definition of $\Phi_{KSW}$\cite{KSW} in events where the thrust
axis is along ${\vec p}_1+{\vec p}_3$ or ${\vec p}_1+{\vec p}_4$.

In situation (b), where there is no $b$-tagging, we label the
jets according to their energy, such that
${E_1\ge E_2\ge E_3\ge E_4}$, and then define the angles by
eqs.~(\ref{def:NR})-(\ref{def:KSW2}) as before.

By considering the polarization of the gluon in $e^+e^-\ar q\bar q g$ and
the final state helicities in the subsequent splitting $g \to gg$ or $g \to
q'\bar{q}'$, one finds that the $e^+e^- \to q \bar{q} gg$ cross section is
concentrated near $\cos\theta_{NR}^*\approx \pm 1$, whereas in $e^+e^- \to
q \bar{q} q' \bar{q}'$, the cross section is largest around
$\cos\theta_{NR}^*\approx 0$. In the case of the Bengtsson-Zerwas angle,
one expects the $g\ar gg$ contribution to be rather flat in the
corresponding distribution if compared with the $g\ar q\bar q$ one, which
generally peaks at $\chi_{BZ}\approx 90^\circ$. The original
K\"orner-Schierholz-Willrodt angle is defined for events for which there
are two jets in both the hemispheres separated by the plane perpendicular
to the thrust axis: it is the angle between the oriented normals to the
plane containing the jets in one hemisphere and to the plane defined by the
two other jets. The advantage of the modified definition $\Phi_{KSW}^*$
adopted here is that it allows us to include the complete $4$jet-sample in
the analysis, without having to discard events with three vectors in the
same hemisphere. In the splitting process $g\ar gg$, the two planes tend to
be parallel, with the two gluons on the same side, i.e.,
$\Phi_{KSW}^{(*)}\approx\pi$, whereas the planes are preferentially
orthogonal for $g\ar q\bar q$. Finally, gluons from the triple gluon vertex
$g\ar gg$ and the second pair of quarks from $g\ar q\bar q$ are expected to
be closer together than gluons from double bremsstrahlung, and this should
be evident by looking at the angle between the two softest jets. More
details on these arguments are given in ref.~\cite{nonAbel}.

The results for the Nachtmann-Reiter angle are shown in figs.~3a~\&~b.
In the case of $b$-tagging (fig.~3a), the distributions are even functions
of $\cos\theta_{NR}^*$, because replacing $\theta_{NR}^*$ by
$\pi-\theta_{NR}^*$ is equivalent to interchanging ${\vec p}_3$ and
${\vec p}_4$, and we do not make any distinction between jets 3 and 4.
The $b\bar{b}gg$-distributions
have peaks at $\cos\theta_{NR}^*=\pm1$, whereas the maxima of the
$b\bar{b}q\bar{q}$ are at $\cos\theta_{NR}^*=0$. We also note that the
$b\bar{b}gg$-distributions in QCD and in the Abelian model are different.
Without $b$-tagging, the distributions of $\cos\theta_{NR}^*$
(fig.~3b) are skewed to the left. The reason for this asymmetry
is kinematical and can be understood by considering events where all four
jets are all close to one common axis (with small angles between them in
order to satisfy the $y$-cuts). Then, by momentum conservation, the most
and the next most energetic jets, 1 and 2, must go in opposite
directions. If we further restrict our attention to events where jets 3
and 4 also go in opposite directions, then energy-ordering and momentum
conservation together imply that jet 3 must be parallel to jet 2, which
gives $\theta_{NR}=\pi$, and cannot be parallel to jet 1, which would give
$\theta_{NR}=0$. Although the $q\bar{q}gg$ and
$q\bar{q}q'\bar{q}'$-distributions are still different, the difference between
QCD and the Abelian model in the $q\bar{q}gg$-distribution is washed away.

In figs.~4a~\&~b, the distributions are shown of the Bengtsson-Zerwas
angle, with and without $b$-tagging, respectively. With $b$-tagging the
distributions are again symmetric, for the same reason as in the case of
Nachtmann-Reiter. At $\chi_{BZ}=\pi/2$, there is a peak in the
$b\bar{b}q\bar{q}$-distributions and a dip in the
$b\bar{b}gg$-distributions. Without $b$-tagging, the
$b\bar{b}gg$-distributions are shifted to lower values of $\chi_{BZ}$,
whereas the $b\bar{b}q\bar{q}$-distributions are shifted slightly towards
higher $\chi_{BZ}$.

The distributions where the rewards for $b$-tagging are largest are
probably those of the angle $\Phi_{KSW}^*$, shown in figs.~5a~\&~b. With
$b$-tagging, the differences between the $b\bar{b}gg$ distributions in QCD
and the Abelian model become even more clear than in $\chi_{BZ}$. In this
respect the modified definition of $\Phi_{KSW}^*$, initially adopted in
order to avoid loss of statistics, turns out to be extremely successful.

Finally, figs.~6a~\&~b show the distribution of $\cos\theta_{34}$. Here we
see, as expected, a tendency for the gluons from the triple gluon vertex,
and the quarks $q\bar{q}$ in $b\bar{b}q\bar{q}$ events, to be closer
together than the gluons in the Abelian model. In fig.~6a, there is also
a large concentration of $b\bar{b}q\bar{q}$ events near $\cos\theta_{34}=-1$.
They come from the region of phase space where the $b$-quarks are relatively
soft, where the cross section is dominated by the diagrams in which the
$Z^0$ is coupled directly to the quarks $q\bar{q}$. That explains why the
peak disappears completely when the jets are energy-ordered (fig.~6b).

We should warn the reader that the value of $y_{cut}$ we have chosen for
using with the Geneva algorithm corresponds to a looser cut than the ones
we use with the other algorithms. This is because, unlike the other jet
defining variables, the definition (\ref{GENEVA}) of $y^G$ does not contain
$s$ explicitly, and therefore, the Geneva algorithm allows the energies of
the partons to be much smaller for a given value of $y_{cut}$. As a result,
we can get very close to the singularities of the matrix elements, where we
expect radiative corrections to be large (figs.~5 and 6). The large peaks
in the cross section also make it more difficult to integrate by VEGAS, as
can be seen from the statistical fluctuations in the distributions of
figs.~3~\&~4. Therefore, a larger value of $y^G_{cut}$ would be needed to
obtain reliable predictions from our tree-level calculation \cite{GENEVA},
but we prefer to keep the value shown just to illustrate what happens.

\section{Photon sample}
\label{subs:photon}

In the literature \cite{workshop,Mattig}, it has been argued that one might
be able to see the non-Abelian structure of QCD by comparing the
distributions of four-jet events with those of events where three jets and
a photon are produced. Since the diagrams of fig.~2a are the same, up to a
constant factor, for the processes (\ref{proc2}) and (\ref{proc3}), any
differences in their distributions must be due to the non-Abelian diagrams
of fig.~2b and the ISR diagrams of fig.~2c. As we shall
show below, the contribution of the latter can be made negligibly small by
applying suitable cuts. If one then assumes that the four-jet events are
predominantly $q\bar{q}gg$ events, as is true in QCD, differences between
the four-jet and the three-jet plus photon distributions can be regarded as
evidence for the non-Abelian contribution
$|{\cal M}_b|^2$ in eq.~(\ref{M2-2q2g}).

What effect would $b$-tagging have on such an analysis? Presumably, the
distinctions between the distributions would become more clear, but
the number of events would be smaller. Moreover, selecting
events with a $b\bar{b}$-pair in the final state increases the relative
number of unwanted four-quark events, since five of the fifteen flavour
combinations $q\bar{q}q'\bar{q}'$, but only one of the five combinations
$q\bar{q}gg$ (where $q,q'=u,d,c,s,b$), contain at least one
$b\bar{b}$-pair. It is possible to reduce this contamination of the
four-jet sample by imposing cuts, at the cost of a further loss of
statistics. The crucial question is, whether, in the end, the event
rates would still be large enough to allow a study of the distributions.

The ISR diagrams are important for photons that are
either soft or nearly collinear with the incoming electrons and positrons.
Therefore, we can eliminate them by imposing cuts on the photon energy
$E_{\gamma}$ and on the angle $\theta_{beam-\gamma}$ between the photon
trajectory and the $e^{\pm}$ beams. Since we wish to compare the sample of
three-jet plus photon events with an equivalent sample of four-jet events,
we must treat both samples on exactly the same footing. In the four-jet
sample we impose the same cuts on all four jets, or at least on the two
non-$b$ jets, in case we select events where two of the jets are tagged as
$b$-jets. This implies that in the three-jet plus photon sample, we must
also impose the same cuts on the jet energies $E_{jet}$ and their angles
with respect to the beams $\theta_{beam-jet}$. In the $b$-tagged case,
we only apply these cuts to the non-$b$ jet.

It turns out that demanding that $|\cos \theta_{beam-\gamma}|<0.9$ and
$E_{\gamma}>10$~GeV is sufficient to remove the effect of the ISR
diagrams. This is illustrated, for example, in fig.~7, where we
have implemented the cut $|\cos \theta_{beam-\gamma}|<0.9$ and plotted the
differential distribution in the photon energy of the three-jet plus photon
cross section twice: taking the ISR diagrams into account and omitting
them. Above our 10 GeV cut on $E_{\gamma}$, the two distributions are the
same. We have checked that, with these cuts, the distributions of other
variables also look the same with and without ISR.

The total cross section for process~(\ref{proc3}) is given in tab.~I
for several values of $y_{cut}$. The effect of the ISR
is never greater than a few percent.

In tab.~II we show the $q\bar{q}gg$ and $q\bar{q}q'\bar{q}'$ components
of the total $e^+e^- \to 4 \mbox{jet}$ cross section. In tab.~IIa, where
we assume two $b$-jets are tagged, we have imposed two additional cuts,
namely $|\cos\theta_{NR}^*|>0.5$ and $\chi_{BZ}<50^{\circ}$ or
$\chi_{BZ}>130^{\circ}$. As can be seen in figs.~3~\&~4, this reduces
the relative size of the $q\bar{q}q'\bar{q}'$ component, making it less
than about 8\% of the total four jet cross section for all values of
$y_{cut}$ shown in the table. We have not imposed these cuts in tab.~IIb,
which shows the results if one does not select $b$-tagged jets,
because there the $q\bar{q}q'\bar{q}'$ component is already quite small
without them.

A comparison between the $e^+e^- \to q\bar{q}gg$ and
$e^+e^- \to q\bar{q}g\gamma$ cross sections is made in fig.~8.
They are displayed as a function of $y_{cut}$ for each of the jet-finding
algorithms. The dotted curves marked ``real'' show the actual
$q\bar{q}g\gamma$ cross section. The dashed curves marked ``renormalised''
show the $q\bar{q}g\gamma$ cross section multiplied by
\be
\frac{  \alpha_{s} C_a^{q\bar qgg}}
     {2 \alpha_{em} e_q^2 C^{q\bar qg\gamma}}.
\ee
Apart from the small ISR effect, this is exactly
the contribution of the QED-like graphs, i.e., the term $|{\cal M}_a|^2$
in eq.~(\ref{M2-2q2g}), to the $e^+e^- \to q\bar{q}gg$ cross section.
The factor of two in the denominator is the symmetry factor needed to
account for the two identical gluons in the $q\bar{q}gg$ final state.
The non-Abelian contribution $|{\cal M}_b|^2$ is the
difference between this ``renormalised'' $q\bar{q}g\gamma$ cross
section and the $q\bar{q}gg$ cross section.

\section{Mass effects}
\label{subs:me}

In this section, we examine the numerical importance of taking the
$b$-quark mass into account exactly, as we have done in all the
calculations we have discussed until now. The $b$-quark mass does not only
affect the total cross section \cite{noiPL,noiNP,noiPr},  but also some of
the angular distributions. In figs.~9, 10 \&
11, we present plots illustrating this for the processes
$e^+e^- \to b \bar{b}gg$, $e^+e^- \to b \bar{b} u \bar{u}$, and
$e^+e^- \to b \bar{b} b \bar{b}$. We do not show, e.g.,
$e^+e^- \to b \bar{b} d \bar{d}$, because the results are quite
similar to those of $e^+e^- \to b \bar{b} u \bar{u}$. In each
case, we compare the distributions obtained for $m_b=5$~GeV with
the ones we would find if we neglected $m_b$.

The differences turn out to be fairly small in the $b \bar{b}gg$-process,
but in the other ones they are quite large, particularly in the
distributions of $\chi_{BZ}$ and $\Phi^*_{KSW}$. This is true both for the
distributions where the $b$-jets are identified, see figs.~10a \& 11a, and
for the distributions where the jets are simply ordered in energy,
figs.~10b and 11b. These mass effects also depend on the particular
jet-finding algorithm used. As an example we plotted the case of the E
scheme, for which the mass effects are larger.

In the equal flavour process, $e^+e^- \to b \bar{b} b \bar{b}$, there are
three different combinations of $b$-(anti)quarks that can be tagged: two
$b$'s, one $b$ and one $\bar{b}$, or two $\bar{b}$'s. That is why fig.~11a
shows two sets of curves, one for the $bb$ case and one for the $b\bar{b}$
case\footnote{In figs.~3a, 4a, 5a \& 6a, we assumed a $b$ and a $\bar{b}$
were tagged. If that assumption is not true, the contribution of the
$e^+e^- \to  b \bar{b} b \bar{b}$ events to those distributions has to be
replaced with a weighted average of the two sets of curves in fig.~11a.
However, since the contribution of the $e^+e^- \to  b \bar{b} b \bar{b}$
events is small, the distributions in figs.~3a,  4a, 5a \& 6a would hardly
change.}. We also note that even in the $b\bar{b}$ case, the distribution
of $\cos\theta_{NR}^*$ is very different from the corresponding
distribution in the unequal flavour process. Due to the gluon propagators
in the last four diagrams of fig.~1, it peaks at $\cos\theta_{NR}^*=\pm1$,
making it look similar to the $\cos\theta_{NR}^*$ distribution of the
$b\bar{b}gg$ process.

\section{Conclusions}
\label{subs:concs}

We have given our results in terms of total and differential cross
sections. To convert them into event rates, they must be multiplied by the
luminosity and the efficiency for tagging two $b$-quarks. We have not done
this because the numbers differ from one experiment to another, and could
still change as the techniques are improved. However, to get an idea of
what one might expect, let us suppose the integrated luminosity is 100
pb$^{-1}$ and the tagging efficiency 50\%. Then there will be roughly
$4\times10^4$ four jet events with two tagged $b$-quarks (using the JADE
scheme with $y_{cut}=0.01$, as in figs.~3-6). This is an order of
magnitude less than the total number of four jet events, but in return we
gain greater power to discriminate between the various terms in the four
jet cross section, particularly between the Abelian, QED-like term and the
non-Abelian term in the two quark, two gluon cross section.

In a realistic analysis, one should take the probability of misidentifying
other particles as $b$-quarks into account. We do not expect $c$-jets
mistagged as $b$-jets to distort the distributions severely, although they
would increase the total number of events, because they would still be
correctly classified as quark jets. It is, of course, important that the
number of gluon jets tagged as $b$'s be as small as possible. The
probability of a $b$-quark being created during the fragmentation of a
gluon or a lighter quark is believed to be negligible \cite{deangelis}.

The number of three jet $\gamma$ events is reduced even more severely by
demanding two tagged $b$-quarks, because of their small electric charge
($-1/3$ instead of $2/3$). With the luminosity and efficiency given above,
and the cuts of fig.~8a, one would expect of the order of 50 events, which
is not enough to study any distributions. So, in the photon sample, it is
better to keep all events.

Finally, we indicated the effects of the $b$-quark mass. Normally, they can
safely be neglected because only a small fraction of the four jet events
contain $b$-quarks, and moreover, most of those are $b\bar{b}gg$ events,
where the effects are small. This is no longer true with $b$-tagging,
because now all events contain at least two massive $b$-quarks, and the
relative number of four quark events is higher, and the mass effects are
especially important when comparing QCD with the Abelian model, where as
many as 30\% of the events are four quark events.

\section*{Acknowledgements}

We are grateful to E.~Maina for indicating to us part of the bibliography
on the subject discussed here, and to E.W.N.~Glover for urging us to study
the quark mass dependence of the angular distributions. We would like to
thank Marco Paganoni, Marco Bigi and Alessandro De Angelis for useful
discussions, and S.M. thanks the Mainz Dept. of Physics for kind
hospitality. J.B.T. is supported by the Graduiertenkolleg
``Teilchenphysik'' in Mainz.

\newpage
\thispagestyle{empty}
\subsection*{Table Captions}
\begin{description}

\item[tab.~I  ]

Cross sections in picobarns of the
processes (a) $e^+e^-\rightarrow b\bar bg\gamma$ (i.e., $b$-tagging)
and (b) $e^+e^-\rightarrow \sum_q q\bar qg\gamma$ (i.e., no $b$-tagging),
with and without ISR,
for three different values of $y_{cut}$ of each jet-finding
algorithm, using the notation $(\sigma_{ISR},\sigma_{noISR};y_{cut})$.

The following additional cuts have been implemented:
$|\cos\theta_{\mbox{\scriptsize beam}-\gamma,g}|<0.9$
and $E_{\gamma,g} >10$ GeV, for case (a), and
$|\cos\theta_{ \mbox{\scriptsize beam}-\gamma,g,q,\bar{q}}|<0.9$
and $E_{\gamma,g,q,\bar{q}} >10$ GeV, for case (b).

\item[tab.~II ]

Cross sections in picobarns of the processes (a) $e^+e^-\rightarrow b\bar
bgg$ ({\it x}) and $e^+e^-\rightarrow \sum_q b\bar bq\bar q$ ({\it y})
(i.e., $b$-tagging) and (b) $e^+e^-\rightarrow \sum_q q\bar qgg$ ({\it x})
and $e^+e^-\rightarrow \sum_q\sum_{q'} q\bar qq'\bar q'$ ({\it y}) (i.e.,
no $b$-tagging), for three different values of $y_{cut}$ ({\it z}) of each
jet-finding algorithm, adopting the notation $(x,y;z)$.

The following additional cuts have been implemented:
$|\cos\theta_{ \mbox{\scriptsize beam}-g}|<0.9$,
$E_g>10$ GeV, $|\cos\theta_{NR}^*|>0.5$ and
$\chi_{BZ}<50^{\circ}$ or $\chi_{BZ}>130^{\circ}$, for case (a), and
$|\cos\theta_{ \mbox{\scriptsize beam}-g,q,\bar{q}}|<0.9$
and $E_{g,q,\bar{q}}>10$ GeV, for case (b).
\end{description}

\vspace*{\fill}

\newpage
\thispagestyle{empty}
\subsection*{Figure Captions}
\begin{description}
\item[fig.~1 ]
Feynman diagrams contributing in lowest order to
$e^+e^-\rightarrow q\bar q q'\bar q'$, where $q^{(')}=u,d,s,c$ and $b$.
If $q\ne q'$ only the first four diagrams contribute.
The internal wavy line represents a photon or a $Z^0$.
The particles are labelled as in eq.~(\ref{proc1}).
\item[fig.~2 ]
 Feynman diagrams contributing in lowest order to
$e^+e^-\rightarrow q\bar qgg$ (a and b) and
$e^+e^-\rightarrow q\bar qg\gamma$
(a and c), where $q=u,d,s,c$ and $b$.
The internal wavy line represents a photon or a $Z^0$, while the external
jagged line represents a gluon or a photon, as appropriate.
The particles are labelled as in eqs.~(\ref{proc2})-(\ref{proc3}).
\item[fig.~3 ]
Distributions in the cosine of the modified Nachtmann-Reiter angle,
$\cos\theta_{NR}^*$,
in (a) $e^+e^-\rightarrow b\bar bgg,\sum_qb\bar b q\bar q$ (i.e., $b$-tagging)
and (b) $e^+e^-\rightarrow \sum_q q\bar qgg,\sum_q\sum_{q'} q\bar q q'\bar q'$
(i.e., no $b$-tagging), for
the various jet-finding algorithms.
\item[fig.~4 ]
 Distributions in
the Bengtsson-Zerwas angle, $\chi_{BZ}$,
in (a) $e^+e^-\rightarrow b\bar bgg,\sum_qb\bar b q\bar q$ (i.e., $b$-tagging)
and (b) $e^+e^-\rightarrow \sum_q q\bar qgg,\sum_q\sum_{q'} q\bar q q'\bar q'$
(i.e., no $b$-tagging), for
the various jet-finding algorithms.
\item[fig.~5 ]
 Distributions in
the modified K\"orner-Schierholz-Willrodt angle, $\Phi_{KSW}^*$,
in (a) $e^+e^-\rightarrow b\bar bgg,\sum_qb\bar b q\bar q$ (i.e., $b$-tagging)
and (b) $e^+e^-\rightarrow \sum_q q\bar qgg,\sum_q\sum_{q'} q\bar q q'\bar q'$
(i.e., no $b$-tagging), for
the various jet-finding algorithms.
\item[fig.~6 ]
 Distributions in the cosine of
the angle between the vectors $\vec p_3$ and $\vec p_4$, $\cos\theta_{34}$,
in (a) $e^+e^-\rightarrow b\bar bgg,\sum_qb\bar b q\bar q$ (i.e., $b$-tagging)
and (b) $e^+e^-\rightarrow \sum_q q\bar qgg,\sum_q\sum_{q'} q\bar q q'\bar q'$
(i.e., no $b$-tagging), for
the various jet-finding algorithms.
\item[fig.~7 ]
Distributions in energy of the photon $E_{\gamma}$
in (a) $e^+e^-\rightarrow b\bar bg\gamma$ (i.e., $b$-tagging), with
$|\cos\theta_{ \mbox{\scriptsize beam}-\gamma,g}|<0.9$ and
$E_{\gamma,g}>1.0$ GeV,
and (b) $e^+e^-\rightarrow \sum_q q\bar qg\gamma$ (i.e., no $b$-tagging),
with $|\cos\theta_{ \mbox{\scriptsize beam}-\gamma,g,q,\bar{q}}|<0.9$ and
$E_{\gamma,g,q,\bar{q}}>1.0$ GeV,
with and without ISR, for the various jet-finding algorithms.
\item[fig.~8 ]
Cross sections of the processes
(a) $e^+e^-\rightarrow b\bar bgg,b\bar bg\gamma$
(the latter both real and renormalized to the
former, see in the text) (i.e., $b$-tagging), with
$|\cos\theta_{ \mbox{\scriptsize beam}-\gamma,g}|<0.9$ and
$E_{\gamma,g}>10$ GeV,
and (b)
$e^+e^-\rightarrow \sum_q q\bar qgg,\sum_q q\bar qg\gamma$
(the latter both real and renormalized to the
former, see in the text) (i.e., no $b$-tagging), with
$|\cos\theta_{\mbox{\scriptsize beam}-\gamma,g,q,\bar{q}}|<0.9$ and
$E_{\gamma,g,q,\bar{q}}>10$ GeV,
as a function of $y_{cut}$ for the various jet-finding
algorithms.
The following angular cuts have been also implemented in case (a):
$|\cos\theta_{NR}^*|>0.5$ and
$\chi_{BZ}<50^{\circ}$ or $\chi_{BZ}>130^{\circ}$.

\item[fig.~9 ]

Mass effects in the angular distributions of the process
$e^+e^- \to b \bar{b} gg$ using the E jet finding algorithm.
In (a) $b$-tagging is assumed, in (b) the jets are energy-ordered.
The curves denoted by $b\bar{b}gg$ are for massive, and
those denoted by $d\bar{d}gg$ for massless $b$-quarks.

\item[fig.~10 ]

Mass effects in the angular distributions of the process
$e^+e^- \to b \bar{b} u \bar{u}$ ($q\neq b$), using the E
jet-finding algorithm.
In (a) $b$-tagging is assumed, in (b) the jets are energy-ordered.
The curves denoted by $b\bar{b}q\bar{q}$ are for massive, and
those denoted by $d\bar{d}q\bar{q}$ for massless $b$-quarks.

\newpage
\thispagestyle{empty}

\item[fig.~11 ]

Mass effects in the angular distributions of the process
$e^+e^- \to b \bar{b} b \bar{b}$, using the E
jet-finding algorithm.
(a), solid and short-dashed lines: one $b$ and one $\bar{b}$ are tagged;
dotted and long-dashed lines: two $b$'s are tagged.
In (b), the jets are energy-ordered.
The curves denoted by $b\bar{b}b\bar{b}$ are for massive, and
those denoted by $d\bar{d}d\bar{d}$ for massless $b$-quarks.

\end{description}

\vfill
\newpage
\thispagestyle{empty}
\begin{table}
\begin{center}
\begin{tabular}{|c|c|c|c|}     \hline
\rule[-0.6cm]{0cm}{1.3cm}
J                                        &
E                                        &
D                                        &
G                                    \\ \hline
\rule[-0.6cm]{0cm}{1.3cm}
$(1.76,1.68;0.01)$                            &
$(2.39,2.28;0.01)$                            &
$(2.72,2.62;0.0015)$                          &
$(2.64,2.54;0.0015)$                            \\ 
\rule[-0.6cm]{0cm}{1.3cm}
$(1.08,1.02;0.02)$                            &
$(1.38,1.32;0.02)$                            &
$(2.26,2.17;0.0030)$                          &
$(2.12,2.03;0.0030)$                            \\ 
\rule[-0.6cm]{0cm}{1.3cm}
$(0.68,0.65;0.03)$                            &
$(0.84,0.80;0.03)$                            &
$(1.92,1.83;0.0045)$                          &
$(1.75,1.67;0.0045)$                            \\ \hline
\end{tabular}
\end{center}
\centerline{\Large Table Ia}
\end{table}

\
\vskip5.0cm

\begin{table}
\begin{center}
\begin{tabular}{|c|c|c|c|}     \hline
\rule[-0.6cm]{0cm}{1.3cm}
J                                        &
E                                        &
D                                        &
G                                    \\ \hline
\rule[-0.6cm]{0cm}{1.3cm}
$(15.08,14.73;0.01)$                            &
$(15.78,15.41;0.01)$                            &
$(31.27,30.70;0.0015)$                          &
$(27.66,27.13;0.0015)$                            \\ 
\rule[-0.6cm]{0cm}{1.3cm}
$(8.49,8.27;0.02)$                            &
$(8.77,8.54;0.02)$                            &
$(22.38,21.91;0.0030)$                          &
$(19.31,18.89;0.0030)$                            \\ 
\rule[-0.6cm]{0cm}{1.3cm}
$(5.21,5.06;0.03)$                            &
$(5.36,5.20;0.03)$                            &
$(17.66,17.28;0.0045)$                          &
$(14.93,14.58;0.0045)$                            \\ \hline
\end{tabular}
\end{center}
\centerline{\Large Table Ib}
\end{table}

\vfill
\newpage
\thispagestyle{empty}
\begin{table}
\begin{center}
\begin{tabular}{|c|c|c|c|}     \hline
\rule[-0.6cm]{0cm}{1.3cm}
J                                        &
E                                        &
D                                        &
G                                    \\ \hline
\rule[-0.6cm]{0cm}{1.3cm}
$(89.60,5.93;0.01)$                            &
$(128.44,10.78;0.01)$                            &
$(165.26,10.66;0.0015)$                          &
$(155.97,10.00;0.0015)$                            \\ 
\rule[-0.6cm]{0cm}{1.3cm}
$(47.51,3.19;0.02)$                            &
$(63.23,5.73;0.02)$                            &
$(126.71,8.11;0.0030)$                          &
$(114.68,7.50;0.0030)$                            \\ 
\rule[-0.6cm]{0cm}{1.3cm}
$(26.70,1.79;0.03)$                            &
$(33.98,2.82;0.03)$                            &
$(101.64,6.45;0.0045)$                          &
$(89.16,5.89;0.0045)$                            \\ \hline
\end{tabular}
\end{center}
\centerline{\Large Table IIa}
\end{table}

\begin{table}
\begin{center}
\begin{tabular}{|c|c|c|c|}     \hline
\rule[-0.6cm]{0cm}{1.3cm}
J                                        &
E                                        &
D                                        &
G                                    \\ \hline
\rule[-0.6cm]{0cm}{1.3cm}
$(761.23,39.08;0.01)$                            &
$(809.96,44.94;0.01)$                            &
$(1618.04,75.85;0.0015)$                          &
$(1410.82,64.72;0.0015)$                            \\ 
\rule[-0.6cm]{0cm}{1.3cm}
$(419.70,23.91;0.02)$                            &
$(440.26,26.95;0.02)$                            &
$(1152.07,58.14;0.0030)$                          &
$(980.17,48.20;0.0030)$                            \\ 
\rule[-0.6cm]{0cm}{1.3cm}
$(254.69,15.60;0.03)$                            &
$(265.77,17.25;0.03)$                            &
$(910.63,48.14;0.0045)$                          &
$(752.54,38.85;0.0045)$                            \\ \hline
\end{tabular}
\end{center}
\centerline{\Large Table IIb}
\end{table}
\vfill

\end{document}